\documentclass[a4paper,fleqn,usenatbib,divpdfm]{mnras}

\usepackage[T1]{fontenc}
\usepackage{ae,aecompl}
\usepackage{longtable}
\usepackage{rotating}
\usepackage{multirow}
\usepackage{diagbox}

\usepackage{graphicx}   
\usepackage{amsmath}    
\usepackage{amssymb}    

\title[Testing CDD in $R_h=ct$ universe]{Testing the Distance-Duality Relation in the $R_h=ct$ Universe}

\author[Hu \& Wang]{
J. Hu $^{1}$ \& F. Y. Wang $^{1,2}$\thanks{E-mail:
fayinwang@nju.edu.cn}
\\
$^1$ School of Astronomy and Space Science, Nanjing University, Nanjing 210093, China\\
$^2$ Key Laboratory of Modern Astronomy and Astrophysics (Nanjing University), Ministry of Education, Nanjing 210093, China\\
}

\date{Accepted XXX. Received YYY; in original form ZZZ}

\begin{document}
\label{firstpage}
\pagerange{\pageref{firstpage}--\pageref{lastpage}}
\maketitle

\begin{abstract}
In this paper, we test the cosmic distance duality (CDD) relation
using the luminosity distances from joint light-curve analysis (JLA)
type Ia supernovae (SNe Ia) sample and angular diameter distance
sample from galaxy clusters. The $R_h=ct$ and $\Lambda$CDM models
are considered. In order to compare the two models, we constrain the
CCD relation and the SNe Ia light-curve parameters simultaneously.
Considering the effects of Hubble constant, we find that $\eta\equiv D_A(1+z)^2/D_L=1$ is valid at the
2$\sigma$ confidence level in both models with $H_0 = 67.8\pm0.9$ km/s/Mpc. However, the CDD relation is valid at
3$\sigma$ confidence level with $H_0 = 73.45\pm1.66$ km/s/Mpc. Using the Akaike Information Criterion (AIC) and
the Bayesian Information Criterion (BIC), we find that the
$\Lambda$CDM model is very strongly preferred over the $R_h=ct$
model with these data sets for the CDD relation test.
\end{abstract}




\section{Introduction}
The cosmic distance duality (CDD) relation \citep{eth33}, also
called Etherington relation, plays an important role in
observational cosmology. It relates the luminosity distance ($D_L$)
to the angular diameter distance ($D_A$) by the following equation
\begin{equation}\label{1}
\frac{{{D_L}}}{{{D_A}}}{(1 + z)^{ - 2}} = 1.
\end{equation}
It is valid for all cosmological models based on Riemannian
geometry. The bases of this relation are that the number of photons
is conservative and the photons travel along the null geodesics in a
Riemannian space time \citep{ell07}. This relation plays a key role
in modern astronomy, especially in galaxy observations
\citep{cmsl07,Manl10}, cosmic microwave background (CMB) radiation
observations \citep{Kom11} and gravitational lensing \citep{ell07}.
Therefore, the validity of this relation has been extensively
studied in the past several years. For example, \cite{baku04} found
a 2$\sigma$ violation of CDD relation using $D_L$ from type Ia
supernovae (SNe Ia) and $D_A$ from FRIIb radio galaxies. The angular
diameter distances from X-ray observations of galaxy clusters also
have been used to probe the CDD relation \citep{Hol10,hlr11,Hol12}.
Similar works have also been done \citep{mzz12,ghj12,Hga12}.
\cite{Rasa16} used CMB anisotropy to test the CDD relation. The CDD
relation is also important to explore the cosmology opacity
\citep{lvxi16,hyw17}. In order to test the CDD relation, the
distances of cosmological sources $D_A$ and $D_L$ must be measured
at the same redshift. In principle, the two kinds of distances
should not be measured from any relationship which depends on
cosmological model. Luminosity distance $D_L$ is usually derived from SNe Ia in a
cosmological model. For instance, \cite{suz12} used the
$\Lambda$CDM, $w$CDM and $ow$CDM models to fit the parameters of
Union2.1 SNe Ia and constrain the cosmological. \cite{bet14} have
done the similar work by the JLA sample. The $D_A$ can be achieved
by the Sunyaev-Zel'dovich effect (SZE) and X-ray observations
\citep{suze72,cafu78}. \cite{bon06} used X-ray data and SZE data to
determine the angular diameter distances of 38 galaxy clusters in
the redshift range $0.14\leq z\leq 0.89$. \cite{uza04} investigated
the possible deviation from the CDD relation by analyzing the
measurements of SZE and X-ray emission data of galaxy clusters. They
constrained the parameter $\eta = 0.89_{ - 0.03}^{ + 0.04}$ at
1$\sigma$ confidence level, which is defined as
\begin{equation}\label{2}
\eta(z)\equiv\frac{{D_A^{cluster}}}{{D_L}}{(1 + z)^2}.
\end{equation}
\cite{hlr11} took more other parameterized forms of $\eta(z)$ and
found no departure from CDD relation with $\eta  = 1$. Taking the
$D_L$ directly from SNe Ia is a feasible method. This method use the
$D_L$ near each galaxy cluster.
For instance, \cite{njj11}, \cite{baku04} \cite{Hol10},
\cite{hlr11}, \cite{cali11} and \cite{mzz12} used this method to
test the CDD relation. However, the value of $D_L$ is dependent on
cosmological model. More recently, \cite{lia16} introduced a new way
to test this relation based on strong gravitational lensing systems
\citep{cao15} and SNe Ia. In their work, they constrain $\eta$ , the
parameters of SNe Ia and the parameter of gravitational system
simultaneously. \cite{hba16} also used the strong gravitational
lensing systems, but they used the $\Lambda$CDM and $w$CDM model to
fit the $D_L$ of JLA sample SNe Ia.

In this work, we use the 38 angular diameter distances $D_A$ from
\cite{bon06} and JLA SNe Ia sample to test the CDD relation in
$\Lambda$CDM and $R_h=ct$ models. The $R_h=ct$ model was proposed in
\cite{mel07}.
In the $R_h=ct$ universe, the luminosity distance $D_L$ is expressed
as
\begin{equation}\label{3}
  D_{L}^{R_h=ct}(z)=\frac{c}{{{H_0}}}(1 + z)\ln(1 + z).
\end{equation}
The factor $c/H_0$ is the current gravitational horizon $R_h(t_0)$.
Therefore $D_L$ can be written as
\begin{equation}\label{4}
  D_{L}^{R_h=ct}(z)=R_h(t_0)(1 + z)\ln(1 + z).
\end{equation}
The model is an FRW cosmology, which is based on the cosmological
principle and Weyl's postulate \citep{mel07,meshl12}. The $R_h=ct$
model describes a universe that expands at a constant rate, rather
than an accelerating universe. \cite{mel16a}  proved that the
Friedmann--Robertson--Walker (FRW) metric must obey the zero active
mass condition $\rho$$+3p = 0$. This condition adds some extra
restrictions in standard cosmology model. So the model may face some
challenges in theory. \cite{lew13} argued that the $R_h=ct$ model
will deviates from its intended evolution with any density of
matter. \cite{klh16} showed that the zero active mass condition
holding at all epochs is not necessary condition for the symmetries
condition of FRW spacetime. The argument contradicted the proof of
\cite{mel16a}. \cite{mel17} argued that \cite{klh16} made a mistake
of canonical transformation of the coordinates. Therefore, the zero
active mass condition must be required for the FRW metric. In terms
of the fitting of observational data, the model also has a lot of
controversy. \cite{bise12} pointed out the problems with the
foundations of the $R_h=ct$ model and found that the consequences
for the evolution of the Universe is disfavored by SNe Ia samples
and Hubble rate data. \cite{mel14} found that the $R_h=ct$ model can
explain the CMB anomalies much better than the $\Lambda$CDM model.
The $R_h=ct$ model can explain the creation of supermassive black
holes in early universe more flexibly \citep{mel13}. \cite{tut16}
used the CMB scale information to test the $\Lambda$CDM and $R_h=ct$
models and found that the $\Lambda$CDM model is very strongly
favored over the $R_h=ct$ model. \cite{shaf15} used the JLA SNe Ia
sample and the BAO data to study the $\Lambda$CDM model and $R_h=ct$
model. They found the $\Lambda$CDM model is favored.

\cite{ngs16} claimed that the evidence for the acceleration is
marginal ($\leq3 \sigma$) using a modified statistical model to
analysis the JLA data set. Later, \cite{Ruha16} have strongly
criticized this method and presented the evidence for acceleration
to be $\sim4.2 \sigma$. \cite{har16} showed that the evidence for
acceleration is 4.56 $\sigma$ with the less flexible modeling of
\cite{ngs16}. \cite{shar16} developed a new method to fit
the (JLA) data set in $\Lambda$CDM and XCDM models. But, the result
is significant different from previous work. \cite{rime16} asserted
that our universe is accelerating at 95\% confidence level and the
coasting universe with a negative curvature is ruled out.
\cite{tut17} found that the power-law model has a weak advantage
than the $\Lambda$CDM model. \cite{yuwa14} found that the $R_h=ct$
model can explain some old objects, including the old quasar APM
08279+5255 and five galaxy clusters. But some galaxy clusters with
old ages can not be accommodated in this model. \cite{jjw15} used
the angular size of galaxy clusters to test the several cosmology
models and found that the data favored the $R_h=ct$ model.
The fitting of the Supernova Legacy Survey (SNLS) sample is
also support the $R_h=ct$ model \citep{jjw15b}. \cite{yuwa15} use
strong gravitational lens data to test $\Lambda$CDM and $R_h=ct$
models. They found that the $R_h=ct$ model is superior to the
$\Lambda$CDM with a likelihood of 74.8 percent using a sample of 12
time-delay lensing systems with Akaike Information Criterion (AIC).
More articles that support the $R_h=ct$ model can be found
in \citep{mefa16, mel16b, mel16c}. So the $R_h=ct$ model is worth of
restudying.

This paper is organized as follows. In next section, we show
the data and the method of parameter fitting in the $R_h=ct$ and
$\Lambda$CDM models. In section 3, the numerical results are shown.
Section 4 gives the conclusions.

\section{Method}
In this work, we explore the CDD relation with the $D_A$ sample from
\cite{bon06}, which consists 38 angular diameter distances of galaxy
clusters. The JLA sample including 740 SNe Ia in the redshift range
$0.01\leq z \leq1.30$ are also used \citep{bet14}. Because the
redshifts of the galaxy clusters are different from the majority of
SNe Ia in the JLA sample. We select a subsample of JLA sample which
contains 38 pairs of SNe Ia to interpolate at the $z_i^{cluster}$
point. The subsample of JLA sample are selected by the following
condition: $z^{SNe}$ satisfies $\Delta z = |z^{SNe} -
z_i^{cluster}|\leq0.005$. The JLA sample and the angular diameter
distance sample are used to fit the CCD relation and the SNe Ia
light-curve parameters simultaneously. The preferred model
must fit both the SNe Ia data and the CDD relation well.

\subsection{Fitting $D_L$ in the $R_h=ct$ universe}
From the phenomenological point of view, the distance modulus of a
SN Ia can be extracted from its light curve, assuming that SNe Ia
with identical color, shape and galactic environment, have on
average the same intrinsic luminosity for all redshifts
\citep{bet14}. This hypothesis is quantified by an empirical linear
relation, yielding a standardized distance $\mu=5\log(D_L/10$pc$)$:
\begin{equation}\label{5}
\mu=m_B^ \star-(M_B-\alpha \times X_1+\beta \times C),
\end{equation}
where $m_B^ \star$ corresponds to the observed peak magnitude in
rest frame $B$ band and $\alpha$, $\beta$, and $M_B$ are nuisance
parameters in the distance estimation. The absolute magnitude is
related to the host stellar mass ($M_{stellar}$) by a simple step
function
\begin{equation}\label{6}
{M_B} = \left\{ \begin{array}{l}
M_B^1\qquad \qquad {\rm{if}} ~~{M_{stellar}}{\rm{ <  1}}{{\rm{0}}^{10}}{M_ \odot }{\rm{      }},\\
M_B^1 + {\Delta _M} \qquad {\rm{otherwise}}.
\end{array} \right.
\end{equation}
The light-curve parameters ($m_B^ \star$, $X_1$ and $C$) are derived
by fitting a model of the SNe Ia spectral sequence to the
photometric data. In our analysis we assume $M_B$, $\Delta_M$,
$\alpha$ and $\beta$ as nuisance parameters.
To determine the value of parameters, we use the maximum likelihood (ML) function \citep{Agostini05,Kim11}
\begin{equation}\label{8}
 \ln(L_{I})=-\frac{1}{2}\sum\limits_{SNe} {(\frac{[\mu (\alpha ,\beta ,M_B) -
 \mu(z; H_0)]^2}{\sigma^2}+\ln(2\pi\sigma^2))},
\end{equation}
where $\sigma^2=\sigma_{lc}^2+\sigma_{ext}^2+\sigma_{sys}^2$ is the
total uncertainty of the SNe Ia distance modulus. \\$\sigma_{lc}
^2$=$V_{m_B}$+$\alpha^2$$V_{x_1}$+$\beta^2$$V_C$+2$\alpha$$V_{m_B,x_1}$$-$2$\beta$$V_{m_B,C}$$-$2$\alpha$$\beta$$V_{x_1,C}$
is the propagated error from the covariance matrix, $V$, of the
light curve fitting. The uncertainties due to host galaxy peculiar
velocities, Galactic extinction corrections, and gravitational
lensing are included in $\sigma_{ext}$. $\sigma_{sys}$ is a
dispersion containing sample-dependent systematic errors that have
not been accounted for and due to the intrinsic variation in SNe Ia
magnitude dispersion \citep{ama10}. We use the value of
$\sigma_{sys}$ calculated by \cite{bet14}, which is not depend on a
specific choice of cosmological model.
In order to see the effect of $H_0$ on the CDD relation, we use $H_0 = 73.45\pm1.66$ km/s/Mpc
\citep{Riess18} and $H_0 = 67.8\pm0.9$ km/s/Mpc \citep{Planck}.

\subsection{Fitting $D_L$ in the $\Lambda$CDM universe}
We use the same method as above to fit the SNe Ia in a
flat $\Lambda$CDM cosmology. The luminosity distance in this model is
\begin{equation}\label{9}
D_{L}^{\Lambda CDM}=\frac{c(1+z)}{H_0} \int\limits_0^z {\frac{{dz}}{{\sqrt {\Omega _m(1 + z)^3 + 1 - \Omega _m} }}},
\end{equation}
$ \Omega _m$ is the  mass density parameter \citep{Hogg00}. Then the expression of  ML fitting is expressed as
\begin{equation}\label{10}
\ln(L_{II})=-\frac{1}{2}\sum\limits_{SNe} {(\frac{{{{[{\mu}(\alpha ,\beta ,M_B)
- {\mu}(z; \Omega _m)]}^2}}}{{{\sigma ^2}}}+\ln(2\pi\sigma^2))}.
\end{equation}

\subsection{Testing the CDD relation}

Equation (2) requires that the each galaxy cluster and the
corresponding SN Ia at the same redshift. Therefore, we use the SNe
Ia subsample mentioned in section 2 to interpolate at the
$z_i^{cluster}$ point accurately. We use the nearest neighbor
interpolation method. The interpolated distance
modulus of the source can be calculated by \citep{liang13}
\begin{equation}\label{11}
\mu_{int}(z)=\mu_i+\frac{z-z_i}{z_{i+1}-z_i}(\mu_{i+1}-\mu_i),
\end{equation}
where $\mu_i$ and $\mu_{i+1}$ are the distance modulus of the SNe Ia
at nearby redshifts $z_i$ and $z_{i+1}$, respectively. The
corresponding uncertainty is
\begin{equation}\label{12}
\sigma_{\mu_{int}}=((\frac{z_{i+1}-z}{z_{i+1}-z_i})^2\sigma^2_{\mu,i}+(\frac{z-z_i}{z_{i+1}-z_i})^2\sigma^2_{\mu,i+1})^\frac{1}{2}.
\end{equation}

Since some SNe Ia locate at the same redshifts, the distance modulus
should be calculated by weighting at the same redshift in the
interpolating procedure.
$\bar{\mu}(z)=\sum(\mu_i/\sigma_{\mu_i}^2)/\sum(1/\sigma_{\mu_i}^2)$,
where $\bar{\mu}(z)$ is the weighted mean distance modulus at the
same redshift $z$ with its error
$\sigma_{\bar{\mu}}=(\sum1/\sigma_{\mu_i}^2)^{-\frac{1}{2}}$\citep{Wang07}.

Next, we test the CDD relation using the interpolated luminosity
distances and $D_A$ of galaxy clusters. We use the form of $\eta(z)$
parametrization \citep{njj11}
\begin{equation}\label{13}
\eta(z)=\eta_1+\eta_2z ,
\end{equation}
where $\eta_1$ and $\eta_2$ are free parameters.
It must be noted that the $D_A$ in equation $\frac{{{D_L}}}{{{D_A}}}{(1 + z)^{ - 2}} = \eta(z)$  is not the given sample of
$D_A^{cluster}$ which acquire from the SZE + X-ray surface
brightness observations technique, but $D_A=D_A^{cluster}/\eta^2$.
Therefore, we can get $\eta(z)\equiv\frac{{D_A^{cluster}}}{{D_L}}{(1
+ z)^2}$. Next, We take the ML fitting to evaluate the most probable values for the parameters, which is
realized by maximizing the ML function
\begin{equation}\label{14}
\ln(L_{CDD})=-\frac{1}{2}\sum\limits_i^{38}( {\frac{{[\eta ({z_i})D_L - D_A^c(1+z)^2]^2}}{{\sigma _{i}^2}}}+\ln(2\pi\sigma_{i}^2)),
\end{equation}
where ${\eta _{obs}}(z) = \frac{D_A^{cluster}}{{D_L}}{(1 +
z)^2}$ and $\sigma _{i}^2 =
\eta^2(z_i)(\sigma_{D_L})^2+(\sigma_{D_A}(1+z_i)^2)^2$. The $D_L=10^{(0.2\mu_{int}-5)}$ and the
$\sigma_{D_L}=\sigma_{\mu_{int}} \ln10/5$ are all related to the coefficients
of SNe Ia light-curve parameters, which can be found in equation (\ref{8}) or equation (\ref{10}). Equation (\ref{14})  provides a
model-independent method to test the CDD relation. The total likelihood functions are
\begin{equation}\label{15}
\ln(L_{R_h=ct})=\ln(L_{CDD})+\ln(L_{I}) ,
\end{equation}
for the $R_h=ct$ model, and
\begin{equation}\label{16}
\ln(L_{\Lambda CDM})=\ln(L_{CDD})+\ln(L_{II}),
\end{equation}
for $\Lambda$CDM model, respectively. Therefore, we can compare the two models according to
the value of maximum likelihood function.

\section{Results}
Figures 1 and 2 show the constraints on $R_h=ct$ model and
$\Lambda$CDM model at 1$\sigma$ and 2$\sigma$ confidence levels with
$H_0=73.45\pm1.66$ km/s/Mpc, respectively. For the $R_h=ct$ model,
$\eta_1 =1.048^{+0.063}_{-0.059}$  and $\eta_2
=0.197^{+0.148}_{-0.150}$ at the 1$\sigma$ confidence level are
obtained. From figure 1, it is obvious that the CDD relation is
valid [$(\eta_1, \eta_2)= (1, 0)$] at 3$\sigma$ confidence level in
the $R_h=ct$ universe. Figure 2 shows the best-fitting results in
the $\Lambda$CDM model. The best-fitting values are $\eta_1
=0.999^{+0.056}_{-0.057}$ and $\eta_2 = 0.181^{+0.144}_{-0.142}$ at
the 1$\sigma$ confidence level. The CDD relation is valid at
3$\sigma$ confidence level in the $\Lambda$CDM model. For $H_0 =
67.8\pm0.9$ km/s/Mpc, figures 3 and 4 show the constraints on
$R_h=ct$ model and $\Lambda$CDM model. $(\eta_1, \eta_2)= (1, 0)$
resides in the 2$\sigma$ confidence level for both models.
Therefore, the value of $H_0$ significantly affects the CDD
relation. In the $R_h=ct$ model, the best fittings are $\eta_1
=0.967^{+0.057}_{-0.055}$  and $\eta_2 =0.182^{+0.142}_{-0.143}$ at
the 1$\sigma$ confidence level. The best-fitting parameters are
$\eta_1 =0.922^{+0.053}_{-0.057}$  and $\eta_2
=0.167^{+0.140}_{-0.135}$ at 1$\sigma$ confidence level in the
$\Lambda$CDM model. The best-fitting value of model parameters are
shown in table 1.

Next, we compare the two cosmological models by using two kinds of
standard information criteria, namely the Akaike Information
Criterion (AIC) \citep{aka1974} and the Bayesian or Schwarz
Information Criterion (BIC) \citep{sch78}. They are defined as
\begin{equation}\label{17}
AIC =  - 2\ln L + 2k ,
\end{equation}
and
\begin{equation}\label{18}
BIC =  - 2\ln L + k\ln N,
\end{equation}
where $L$ is the value of the maximum likelihood function which
defined in equation(\ref{8}) and equation(\ref{10}), $k$ is the
number of model parameters and $N$ is the total number of data
points used in the statistical analysis. To compare the two models,
$\Delta X = {X_M} - {X_{\Lambda CDM}}$ (where X = AIC or BIC) is
used. As a general rule, if $\Delta X \leq 2$, the candidate model
has substantial support with respect to the reference model. if $4
\leq \Delta X \leq7$, it indicates that the concerned model is less
supported with respect to the reference model. If $\Delta X \geq10$,
the candidate model has no observational support \citep{nun17}. For
the $R_h=ct$ model, we have seven model parameters. For the flat
$\Lambda$CDM model, the number of model parameters is eight. The
value of AIC and BIC for the two models are shown in table 2. From
this table, we can see that the $\Lambda$CDM model is very strongly
preferred over the $R_h=ct$ model with these data sets for this
test. We also estimate the degree of rejection. For the model
$\mathcal{M}_\alpha(\alpha=1,2$, i.e. $R_h=ct$ and $\Lambda$CDM
respectively), the likelihood can be written as
\begin{equation}
\mathcal{P}(\mathcal{M}_{\alpha})=\frac{{\rm{exp}}(-{\rm AIC}_{\alpha}/2)}{{\rm{exp}}(-{\rm AIC}_{1}/2)+{\rm{exp}}(-{\rm AIC}_{2}/2)}.
\end{equation}
From the value of AIC and BIC in table 2, we find that the
likelihood of $\Lambda$CDM being correct is almost $100\%$. However, it should be noted that
this conclusion is only for the CDD relation test. For some other tests, these two models are comparable \citep{yuwa15,jjw15b,mefa16, mel16b, mel16c}.

\section{Conclusions}
In this paper, we use the JLA SNe Ia sample and the galaxy cluster
sample to test the CDD relation in the $R_h=ct$ universe. As a
reference, we also test this relation in the $\Lambda$CDM model. The
$D_A$ of galaxy clusters are obtained from SZE and X-ray surface
brightness. The CDD relation and the parameters of SNe Ia
light-curve are fitted simultaneously with the galaxy cluster sample
and the JLA sample. The preferred model must fit both the
SNe Ia data and the CDD relation well. The crucial aspect is that
the SNe Ia sub-samples are carefully chosen to have approximately
equal redshifts of the galaxy clusters ($\Delta z <0.005$).
The parameter $\eta(z)$ is parameterized as $\eta(z)=\eta_1+\eta_2z$.
We study the effect of $H_0$ on the CDD relation.
The results show that the CDD relation is valid at the $2\sigma$ confidence level for $H_0 = 67.8\pm0.9$ km/s/Mpc.
If a large $H_0 = 73.45\pm1.66$ km/s/Mpc is considered, the CDD relation is only valid at $3\sigma$ confidence level.
Comparing the fittings of the two models by the information criteria
of AIC and BIC, we find that the $\Lambda$CDM model is very strongly
preferred over the $R_h=ct$ model with these data sets for this
test.

\section*{Acknowledgements}
We thank the anonymous referee for constructive comments. This work
is supported by the National Basic Research Program of China (973
Program, grant No. 2014CB845800) and the National Natural Science
Foundation of China (grants 11422325 and 11373022), and the
Excellent Youth Foundation of Jiangsu Province (BK20140016).





\begin{figure*}
  \centering
  \includegraphics[scale=0.8,angle=0]{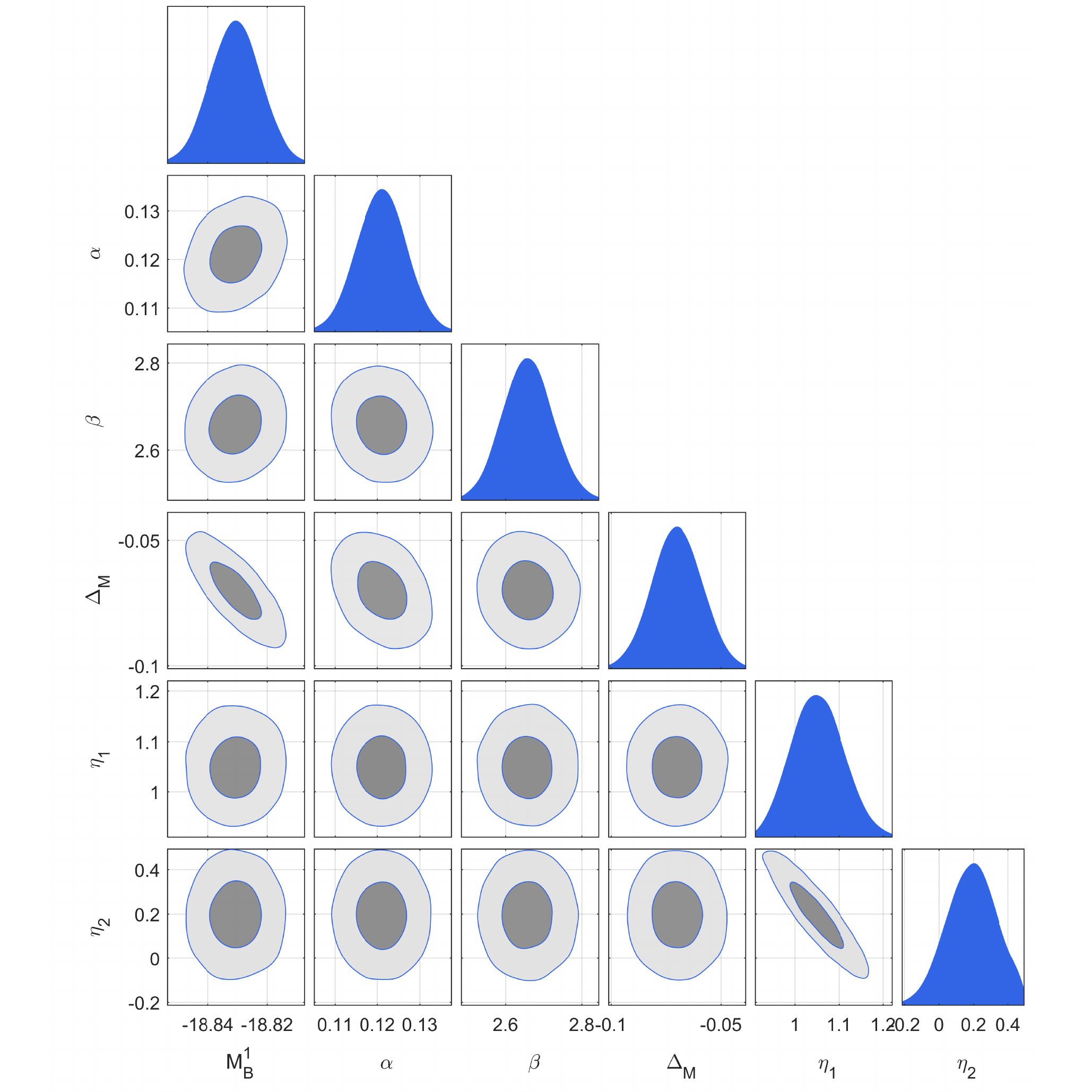}\\
  \caption{In the $R_h=ct$ universe, the 2-D regions and 1-D marginalized distributions with 1$\sigma$ and
2$\sigma$ contours for the parameters $M_B^1$, $\alpha$,
$\beta$, $\Delta_M$, $\eta_1$ and $\eta_2$ using JLA sample and galaxy cluster sample with $H_0=73.45\pm1.66$ km/s/Mpc.}
\end{figure*}

\newpage
\begin{figure*}
  \centering
  \includegraphics[scale=0.8,angle=0]{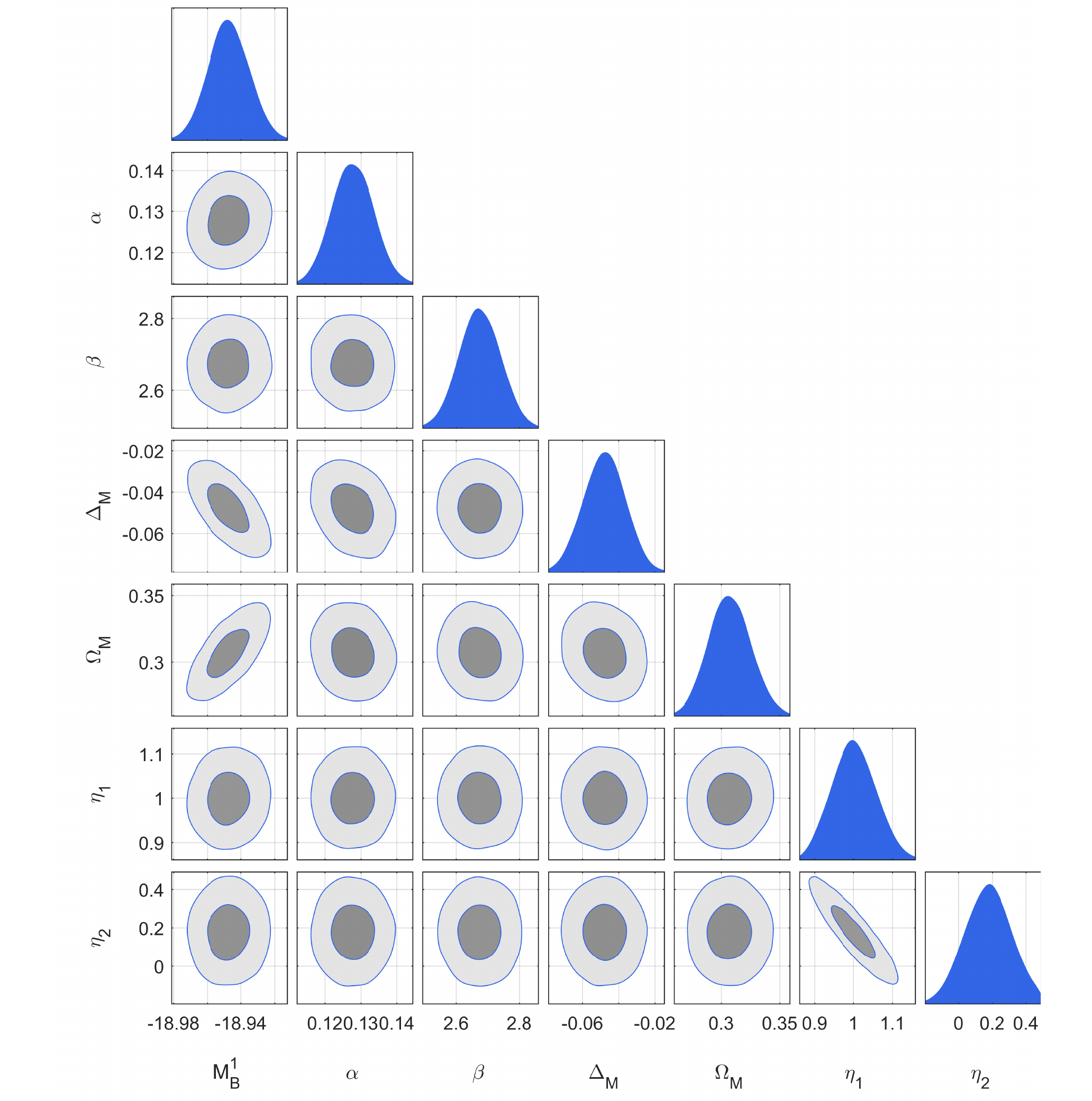}\\
  \caption{In the $\Lambda$CDM model, the 2-D regions and 1-D marginalized distributions with 1$\sigma$ and
2$\sigma$ contours for the parameters $M_B^1$, $\alpha$,
$\beta$, $\Delta_M$, $\Omega_M$, $\eta_1$ and $\eta_2$ using JLA sample and galaxy cluster sample with $H_0=73.45\pm 1.66$ km/s/Mpc.}
\end{figure*}

\begin{figure*}
  \centering
  \includegraphics[scale=0.8,angle=0]{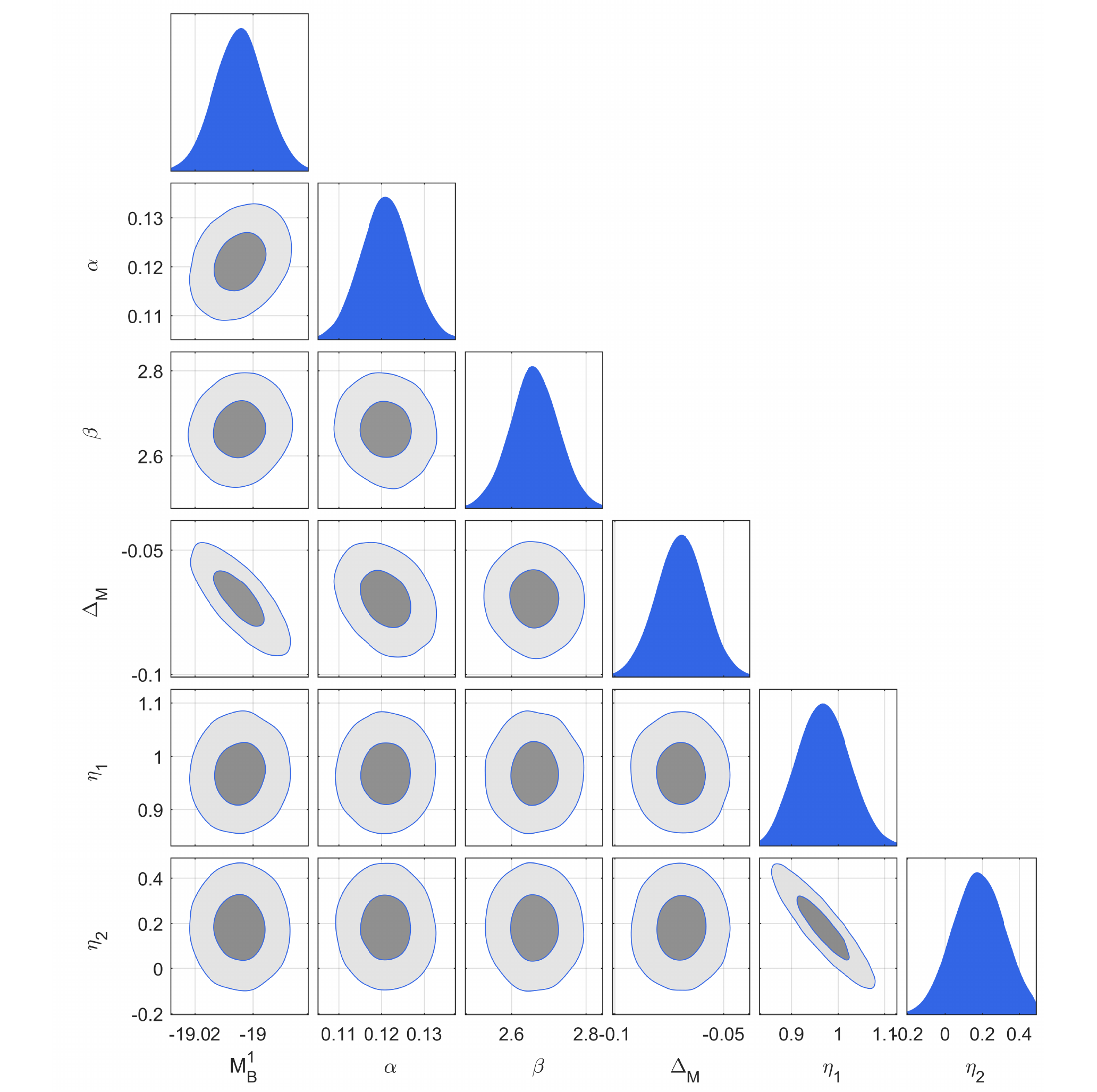}\\
  \caption{In the $R_h=ct$ universe, the 2-D regions and 1-D marginalized distributions with 1$\sigma$ and
2$\sigma$ contours for the parameters $M_B^1$, $\alpha$,
$\beta$, $\Delta_M$, $\eta_1$ and $\eta_2$ using JLA sample and galaxy cluster sample with $H_0=67.8\pm0.9$ km/s/Mpc.}
\end{figure*}

\newpage
\begin{figure*}
  \centering
  \includegraphics[scale=0.8,angle=0]{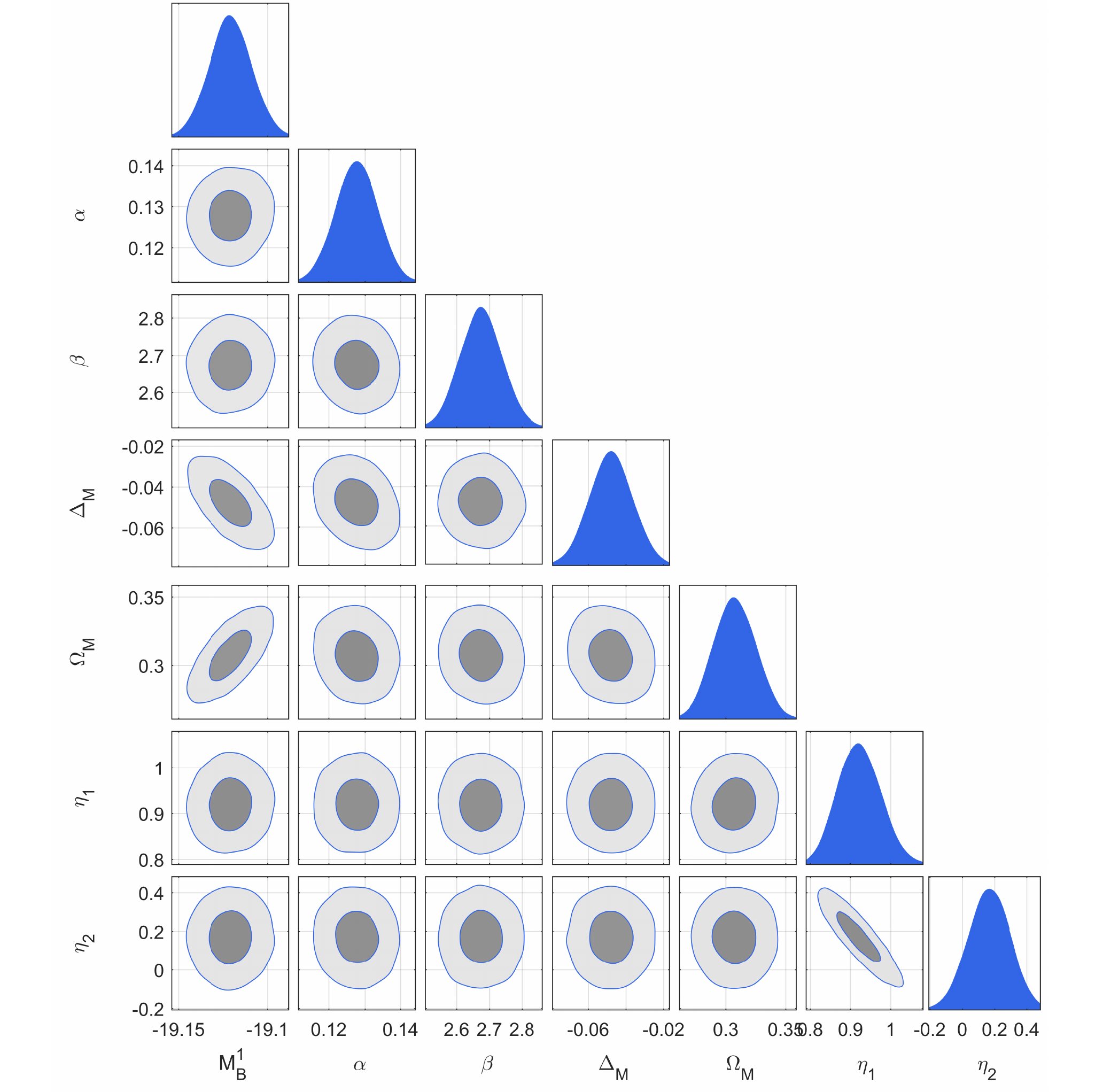}\\
  \caption{In the $\Lambda$CDM model, the 2-D regions and 1-D marginalized distributions with 1$\sigma$ and
2$\sigma$ contours for the parameters $M_B^1$, $\alpha$,
$\beta$, $\Delta_M$, $\Omega_M$, $\eta_1$ and $\eta_2$ using JLA sample and galaxy cluster sample with $H_0=67.8\pm0.9$ km/s/Mpc.}
\end{figure*}

\clearpage
\begin{table*}
\caption{Constraints on the coefficients of light-curve parameters
and cosmological parameters at the 1$\sigma$ confidence levels in
$R_h=ct$ and $\Lambda$CDM models with $H_0 = 67.8\pm0.9$ km/s/Mpc.} \scalebox{0.9}{
\begin{tabular}{|c|l|l|}
\hline
         \diagbox[width=7em,trim=l]{parameters}{model}  &             $\Lambda$CDM  &$R_h=ct$ \\
\hline
        $\alpha$  &  $0.128^{+0.007}_{-0.007}$ & $0.121^{+0.006}_{-0.007}$ \\
\hline
        $\beta$  &        $2.673^{+0.074}_{-0.065}$&        $2.658^{+0.075}_{-0.065}$\\
\hline
        $M^1_B$  &        $-19.121^{+0.012}_{-0.012}$   &        $-19.005^{+0.009}_{-0.009}$    \\
\hline
        $\Delta_M$ &        $-0.048^{+0.014}_{-0.012}$   &        $-0.070^{+0.013}_{-0.011}$    \\
\hline
        $\Omega_M$ &        $0.307^{+0.020}_{-0.019}$ &        $\ldots$ \\

\hline
        $\eta_1$ &         $0.922^{+0.053}_{-0.057}$&        $0.967^{+0.057}_{-0.055}$       \\
\hline
        $\eta_2$ &         $0.167^{+0.140}_{-0.135}$&        $0.182^{+0.142}_{-0.143}$      \\
\hline
        $-2\ln(L)$ &       -62.127&        59.549      \\
\hline
\end{tabular}}
\end{table*}

\begin{table*}
\caption{The values of AIC and BIC.}
\scalebox{1.1}{
\begin{tabular}{|c|l|l|}
\hline
         \diagbox[width=14em,trim=l]{information criteria}{model}  &      $\Lambda$CDM  &      $R_h=ct$       \\
\hline
        AIC &        -46.127 &          73.549           \\
\hline
        BIC &       -8.128&          106.799          \\
\hline
\end{tabular}}
\end{table*}

\bsp    
\label{lastpage}
\end{document}